\documentclass{article} 
\usepackage[preprint]{colm2026_conference}

\usepackage{fancyhdr}

\makeatletter
\usepackage{graphicx}
\usepackage{fancyhdr}

\makeatother

\usepackage{microtype}
\usepackage{hyperref}
\usepackage{url}
\usepackage{booktabs}
\usepackage{tabularx}
\usepackage{amsmath}
\usepackage{multirow}
\usepackage{pifont}
\usepackage{amssymb}
\usepackage{makecell}
\usepackage{adjustbox}
\usepackage{pifont}

\newcolumntype{C}[1]{>{\centering\arraybackslash}p{#1}}
\newcolumntype{Y}{>{\centering\arraybackslash}X}
\newcolumntype{L}[1]{>{\raggedright\arraybackslash}m{#1}}

\usepackage{lineno}
\usepackage{listings}
\usepackage{gradient-text}

\definecolor{darkblue}{rgb}{0, 0, 0.5}
\hypersetup{colorlinks=true, citecolor=darkblue, linkcolor=darkblue, urlcolor=darkblue}

\newcommand{\modelgradient}{\gradientRGB{1D-Bench}{236,47,75}{0,159,255}}

\title{\modelgradient: A Benchmark for Iterative UI Code Generation with Visual Feedback in Real-World}

\author{Qiao Xu$^{\dagger\ddagger}$, Yipeng Yu$^{\dagger}$$\textsuperscript{\ding{41}}$, Chengxiao Feng, Xu Liu\\
Taobao \& Tmall Group of Alibaba\\
\{cunyun.xq, linxin.yyp\}@alibaba-inc.com\\
$^{\dagger}$Equal contribution, \textsuperscript{\ding{41}}Corresponding author\\
$^{\ddagger}$Work done during internship at Alibaba\\}

\begin{document}
\maketitle

\begin{abstract}
Design-to-code translates high-fidelity UI designs into executable front-end implementations, but progress remains hard to compare due to inconsistent datasets, toolchains, and evaluation protocols. We introduce 1D-Bench, a benchmark grounded in real e-commerce workflows, where each instance provides a reference rendering and an exported intermediate representation that may contain extraction errors. 1D is short for one day, representing the efficient completion of design-to-code tasks in less than one day. Models take both as input, using the intermediate representation as structural cues while being evaluated against the reference rendering, which tests robustness to intermediate representation defects rather than literal adherence.

1D-Bench requires generating an executable React codebase under a fixed toolchain with an explicit component hierarchy, and defines a multi-round setting in which models iteratively apply component-level edits using execution feedback. Experiments on commercial and open-weight multimodal models show that iterative editing generally improves final performance by increasing rendering success and often improving visual similarity. We further conduct a pilot study on post-training with synthetic repair trajectories and reinforcement learning based editing, and observe limited and unstable gains that may stem from sparse terminal rewards and high-variance file-level updates. The data and scripts used in this study are available in an anonymized repository at \href{https://anonymous.4open.science/r/d2c-benchmark-A9C4/}{https://anonymous.4open.science/r/d2c-benchmark-A9C4/}.
\end{abstract}

\section{Introduction}

Generating user interface code from design drafts is an important problem spanning HCI and software engineering~\citep{Beyond-Automation}. In practice, designers produce high-fidelity drafts in tools such as Sketch and Figma~\citep{PromptInfuser}, and engineers manually translate them into front-end code. This workflow is costly and error-prone, as small deviations in hierarchy, spacing, or constraints can cause visible inconsistencies and increased maintenance~\citep{yang2025ui2codetextnvisuallanguagemodel,MLLM-Based-UI2Code}.

Recent multimodal large language models have advanced UI code synthesis~\citep{DeclarUI,UICopilot,Divide-and-Conquer,liang2025waffle}. However, many benchmarks rely on synthetic or crawled data and use heterogeneous output targets and toolchains, limiting comparability and alignment with production. Real-world e-commerce UIs further involve deeply nested layouts and diverse constraints that remain challenging.

We present 1D-Bench, grounded in e-commerce development practices, with a standardized evaluation protocol. Each instance includes a reference rendering and an exported design IR that may contain extraction errors~\citep{lu2024omniparserpurevisionbased,Ferret-UI,li2023spotlightmobileuiunderstanding}. Models receive the IR as structural cues while being evaluated against the reference rendering, which tests robustness to IR defects. 1D-Bench requires generating an executable React codebase under a fixed toolchain with an explicit component hierarchy, rather than isolated HTML.

We also define a multi-round setting where models iteratively apply component-level edits using execution feedback. In addition, we conduct an exploratory pilot post-training study using synthetic repair trajectories and reinforcement learning based editing~\citep{jiang2025verltool,wang2025aiagenticprogrammingsurvey}. The observed gains are limited, and we discuss factors that may contribute.

Our contributions are as follows. (1) We introduce 1D-Bench, derived from production e-commerce workflows, that evaluates executable React generation under a fixed toolchain using both a reference rendering and an imperfect exported IR as input. (2) We define a multi-round protocol with execution feedback for iterative editing. (3) We report an exploratory pilot study on post-training for multi-round generation and discuss factors that may limit improvements.

\section{Related Works}
\label{sec:Related Works}

\subsection{Design-to-Code}
With the advancements in the programming capabilities of code LLMs, the design-to-code (D2C) task has garnered renewed interest from researchers in both academia and industry~\citep{Beyond-Automation}. 
Feng used LLMs to generate mid-fidelity wireframes in the UI design process~\citep{feng2023designinglanguagewireframingui}. Petridis developed a Figma plugin that enables designers to author LLM-infused mock-ups~\citep{PromptInfuser}.
Zhou proposed an automated approach that synergizes computer vision, MLLMs, and iterative compiler-driven optimization to generate and refine declarative UI code from designs~\citep{DeclarUI}. Wu used a code compiler and a pre-trained VLM to finetune LLMs to generate UI code from user-provided textual descriptions~\citep{uicoder}. Li used LLMs to detect UI design smells and explain each violation of specific design guidelines in natural language~\citep{electronics13163127}.
Wen presented a document-guided, script-based, end-to-end system named AutoDroid-V2 to support mobile task automation using on-device SLMs~\citep{AutoDroid-V2}. Chen proposed a hierarchy-aware and vision-guided self-correcting approach for generating high-quality UI code from design mockups~\citep{chen2025designcoderhierarchyawareselfcorrectingui}. Wan proposed a divide-and-conquer approach to automate the translation of webpage design to UI code based on MLLMs~\citep{Divide-and-Conquer}. Gui generated UI code from webpage designs by coarse DOM tree prediction and fine-grained code synthesis~\citep{UICopilot}. Xu built a web-based vLLM-based agentic framework for interactive and verifiable UI-to-code generation~\citep{xu2025webviawebbasedvisionlanguageagentic}. Liang proposed a fine-tuning pipeline for UI-to-HTML code generation based on transformer-based MLLMs~\citep{liang2025waffle}. 
However, LLMs struggle to consistently generate UI code that compiles and produces visually relevant designs, and lack fine-grained editing capabilities. One reason is that base vLLMs are not trained on large-scale D2C related data, leading to insufficient coding and visual comprehension abilities. Another reason is that existing training paradigms typically rely on single-round interactions and therefore do not adequately support D2C tasks that require iterative interactions with multiple turns~\citep{yang2025ui2codetextnvisuallanguagemodel,MLLM-Based-UI2Code}.

\subsection{Agentic Learning}
Agentic learning aims to identify effective training methods that enable LLM-based agents to learn from interaction processes and their associated feedback, thereby enhancing their ability to solve real-world problems. The distinction between agent learning and LLM learning lies mainly in two aspects~\citep{zhu2025agenticreinforcementlearningrealworld,zhang2025landscapeagenticreinforcementlearning,zhao2025llmbasedagenticreasoningframeworks,huangetal2025seal,MHSNet25}. First, agent learning involves not only processing text tokens but also handling rich and dynamic contextual environmental information. Second, it requires multi-round interactions with the environment through external tools and API calls~\citep{liu2025toolace,lin2024hammer}. In the area of AI programming, the agentic training paradigm has been widely adopted in foundational LLMs, yet its application to concrete programming tasks remains relatively uncommon~\citep{wang2025aiagenticprogrammingsurvey}. KAT-Coder adopted agentic RL to achieve efficient multi-trajectory optimization, enhanced exploration, and robust policy diversity~\citep{zhan2025katcodertechnicalreport}. GLM-4.5 was trained on agent data in the mid-training stage to increase its capabilities in agentic tasks and coding~\citep{5team2025glm45agenticreasoningcoding}. Kimi K2 synthesized large-scale agentic data and used agentic RL in the post-training~\citep{kimiteam2025kimik2openagentic}. Qwen3 was also trained on agentic tasks in the post-training~\citep{yang2025qwen3technicalreport}. CWM released by Meta is an open-weights LLM for research on code generation with world models~\citep{faircodegenteam2025cwmopenweightsllmresearch}.
Beyond base LLMs, Yu proposed AWorld to orchestrate the training recipe for agentic AI~\citep{yu2025aworldorchestratingtrainingrecipe}. Jiang introduced VERLTOOL to address key limitations of agentic RL with tool use in model training~\citep{jiang2025verltool}. Xiao's work demonstrated the ``less is more'' paradigm for intelligent agency by using only 78 carefully designed training samples~\citep{xiao2025limiagency}. Xiao built a dataset and benchmark of code aesthetic and introduced an agentic reward framework for code generation~\citep{xiao2025codeaestheticsagenticreward}. Majgaonkar analysed trajectories from three state-of-the-art code agents (OpenHands, SWE-agent, and Prometheus) on the SWE-Bench benchmark~\citep{majgaonkar2025understandingcodeagentbehaviour}. Li introduced RepoSearch-R1, a novel agentic RL framework driven by Monte-carlo tree search for repository-level software engineering tasks~\citep{li2025empoweringrepoqaagentbasedreinforcement}. It can be observed that agentic learning is still in its early exploratory phase, with very few studies specifically applying it to D2C tasks, and training data for agentic learning in D2C remain extremely scarce.

\section{1D-Bench}
\subsection{Data Construction}
\label{sec:data_preparation}

\begin{figure*}[!t]
    \centering
    \includegraphics[width=1\linewidth]{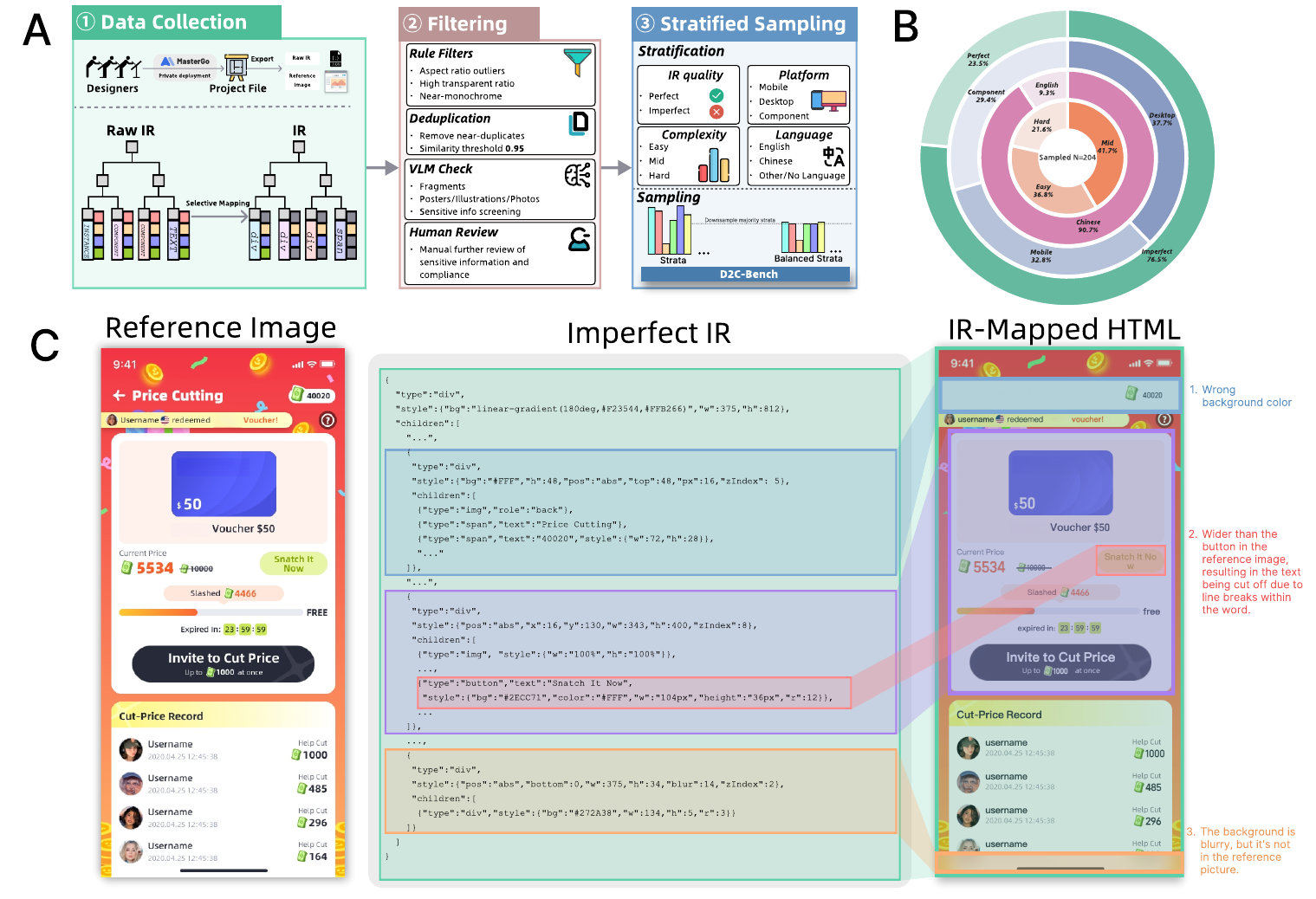}
    \caption{(A) Dataset construction pipeline. (B) Dataset distribution. (C) One dataset example showing the reference UI image and its exported intermediate representation (IR). The IR may contain extraction defects, which propagate to the IR-mapped HTML rendering. }
    \label{fig:data_collection}
\end{figure*}

1D-Bench is derived from an internal static D2C platform used in our organization. For each MasterGo design file, the platform exports an IR and a reference rendering from the same specification. We collect paired IRs and reference images from employee uploads, apply automated filtering, and manually review the retained instances to remove incomplete, low-quality, duplicate, non-UI, or non-compliant records (Appendix~\ref{sec:Implementation of Data Construction}). This yields 984 instances from 5,856 candidates (Figure~\ref{fig:data_collection}A).

We construct the evaluation set of size 204 via stratified sampling over IR reliability, platform type, primary language, and IR size. The IR is provided as an auxiliary, potentially noisy input, while the reference rendering defines the target. The task is to generate an executable React project that matches the reference (Figure~\ref{fig:data_collection}A-B). Figure~\ref{fig:data_collection}C shows an instance.

\subsection{Task Definition}
\label{sec:task_definition}

\begin{figure*}
    \centering
    \includegraphics[width=1\linewidth]{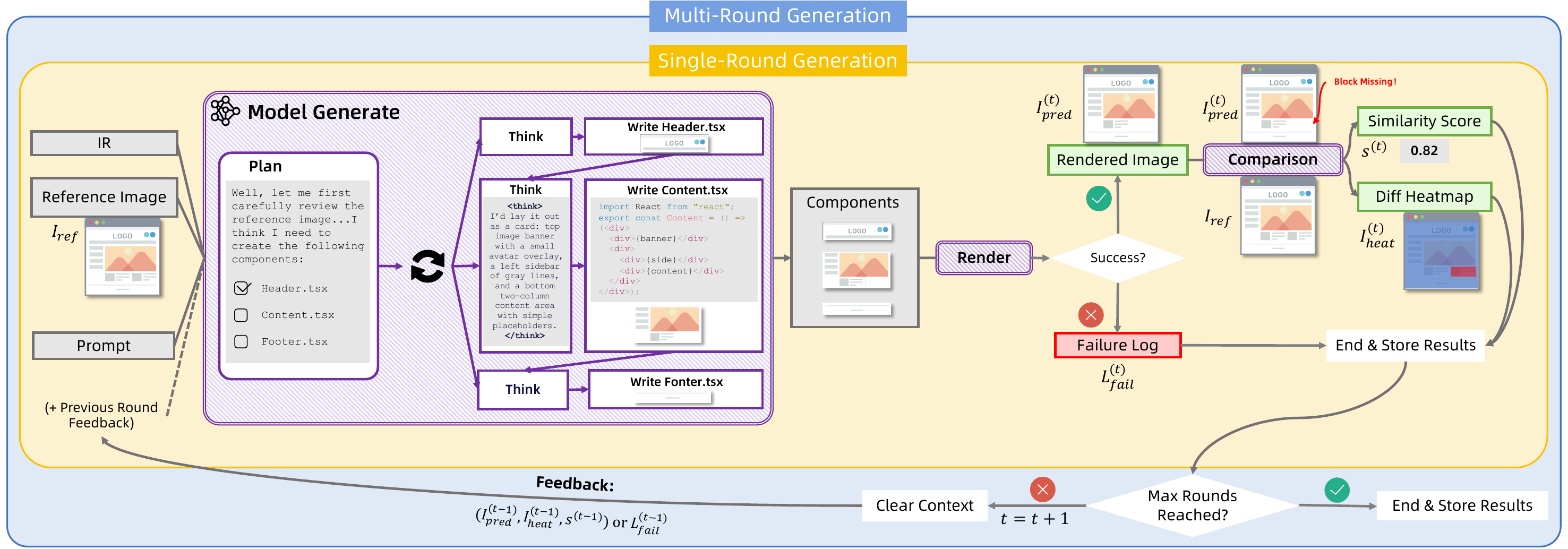}
    \caption{Task definition for single-round and multi-round generation.}
    \label{fig:task_definition}
\end{figure*}

\begin{table*}[t]
  \centering
  \small
  \setlength{\tabcolsep}{2.5pt}
  \renewcommand{\arraystretch}{1.05}

  \begin{adjustbox}{max width=\textwidth}
  \begin{tabular}{l c c c c c c}
    \toprule
     & \textbf{Size} & \textbf{Input} & \textbf{Output} & \textbf{Multi-Round} & \textbf{Type} & \textbf{Real Source}\\
    \midrule
    IW-Bench~\citep{guo2025iw}   & 1200 & Image & HTML  & \ding{55} & Synthetic & -- \\
    pix2code~\citep{beltramelli2018pix2code}   & 1742 & Image & HTML  & \ding{55} & Synthetic & -- \\
    WebCode2M~\citep{gui2025webcodem}  & 768  & Image & HTML  & \ding{55} & Real      & Crawl \\
    CC-HARD~\citep{LaTCoder}    & 128  & Image & HTML  & \ding{55} & Real      & Crawl \\
    MRWeb~\citep{wan2024mrwebexplorationgeneratingmultipage} & 500 & Image & HTML  & \ding{51} & Mixed     & Crawl \\
    FullFront~\citep{sun2025fullfrontbenchmarkingmllmsfrontend} & 400 & Image & HTML  & \ding{55} & Synthetic & -- \\
    Flame-React-Eval~\citep{ge2025advancingvisionlanguagemodelsfrontend} & 109 & Image & React & \ding{55} & Real & Hand-designed \\
    Web2Code~\citep{Web2Code24}   & 5990 & Image & HTML  & \ding{55} & Mixed     & Crawl \\
    Sketch2Code~\citep{li-etal-2025-sketch2code} & 731  & Image & HTML  & \ding{55} & Real      & Hand-designed \\
    Design2Code~\citep{si-etal-2025-design2code} & 484  & Image & HTML  & \ding{55} & Real      & Crawl \\
    \textbf{1D-Bench} & \textbf{204} & \textbf{Image + Metadata} & \textbf{React} & \textbf{\ding{51}} & \textbf{Real} & \textbf{Industry} \\
    \bottomrule
  \end{tabular}
  \end{adjustbox}

  \caption{Public design-to-code benchmarks.}
  \label{tab:public-d2c-benchmarks}
\end{table*}

1D-Bench evaluates executable design-to-code generation given an imperfect exported IR and a reference image that defines the target rendering. For each instance, the model writes a runnable React project under a fixed toolchain. An execution harness builds and renders the project in a standardized viewport and compares the screenshot $I_{\text{pred}}$ with $I_{\text{ref}}$ (Section~\ref{sec:metrics}). Inputs are the exported IR, the reference image, and a pre-initialized workspace with a fixed React scaffold and pinned dependencies (Appendix~\ref{sec:Implementation of Benchmarking}). The MLLMs edit the workspace by creating or overwriting files via a WriteTool-only interface, and the final workspace file tree is evaluated. Specifically, the tasks can be divided into single-round generation and multi-round generation (see Figure \ref{fig:task_definition}).
\begin{itemize}
\item \textbf{Single-Round Generation.}
The model writes the full codebase in one pass. Instances that fail to build or render are counted as rendering failures and receive no similarity score. We report mean similarity over successful renders and the rendering success rate.
\item \textbf{Multi-Round Generation.}
The model iteratively edits the same workspace for up to a fixed number of rounds. After each round $t$, the harness attempts to build and render. On success, it returns $I_{\text{pred}}^{(t)}$, a diff heatmap $I_{\text{heat}}^{(t)}$, and a similarity score $s^{(t)}$; on failure, it returns runtime logs $L_{\text{fail}}^{(t)}$. The process stops at the round limit or when $s^{(t)}$ reaches a preset threshold. We track rendering success across rounds and report final-round results following Section~\ref{sec:benchmarking}.
\end{itemize}

Table~\ref{tab:public-d2c-benchmarks} summarizes public D2C benchmarks. Compared with prior work, 1D-Bench supports multi-round evaluation, uses IR-guided inputs, targets executable React projects under a fixed toolchain, and is sourced from industrial e-commerce data.

\subsection{Metrics}
\label{sec:metrics}

1D-Bench follows an execution-based protocol: the generated React project is built and rendered in a controlled environment, and the resulting screenshot is compared with the reference image. We report (i) visual similarity on successfully rendered instances and (ii) rendering success rate to capture executability. Implementation details are provided in Appendix~\ref{sec:Implementation of Metrics}.

\paragraph{Visual similarity.}
Given the rendered screenshot $I_{\text{pred}}$ and the reference image $I_{\text{ref}}$, we compute a similarity score $S\in[0,1]$ using a composite metric. The metric is primarily based on LPIPS, with auxiliary structural signals (e.g., SSIM) and text/layout cues to better reflect both perceptual fidelity and layout completeness. We report the mean similarity over successfully rendered instances. In the multi-round setting, we use the similarity score from the final round if it renders; otherwise the instance receives no similarity score.

\paragraph{Rendering success rate.}
We record a binary outcome per instance: a run is successful if the project builds and produces a valid screenshot within a fixed timeout, and unsuccessful otherwise. We report the fraction of successful runs over all test instances. In the multi-round setting, failure logs are returned to the model as feedback.

\paragraph{Final reported score.}
To jointly reflect fidelity and executability, we report
\[
\textit{FinalScore}=\bar{S}\cdot \textit{RSR},
\]
where $\bar{S}=\mathbb{E}\!\left[S(I_{\text{pred}}, I_{\text{ref}})\mid \text{render succeeds}\right]$ is the mean similarity over successfully rendered instances, and $\textit{RSR}$ is the rendering success rate. This design prevents methods that overfit visual similarity while frequently failing to build or render from being over-credited.

We also validate metric–human alignment with a small preference study (Figure~\ref{fig:Benchmarking}A): on a 50-instance subset, we generate paired renderings \((R1,R2)\) via different IR perturbations and ask 20 annotators to rank the pair. Human preference correlates with the metric score difference \(\Delta S=S(R1)-S(R2)\).

\section{Benchmarking}
\label{sec:benchmarking}

\subsection{Models}
\label{sec:models_and_baselines}

We evaluate a set of commercial and open-weight multimodal models under a unified protocol, including GPT-5.2, Gemini 3 Pro, Claude Sonnet 4.5, Qwen3-VL-235B-A22B-Instruct~\citep{Qwen3-VL}, and GLM-4.6V~\citep{vteam2025glm45vglm41vthinkingversatilemultimodal}. All models are tested with the same inputs (exported IR and reference image) and the same execution harness (Section~\ref{sec:metrics}). We provide an identical React scaffold with pinned dependencies and a shared WriteTool. Implementation details are provided in Appendix~\ref{sec:Implementation of Benchmarking}.

\subsection{Main Results}
\label{sec:main_results}

\begin{figure*}
    \centering
    \includegraphics[width=1\linewidth]{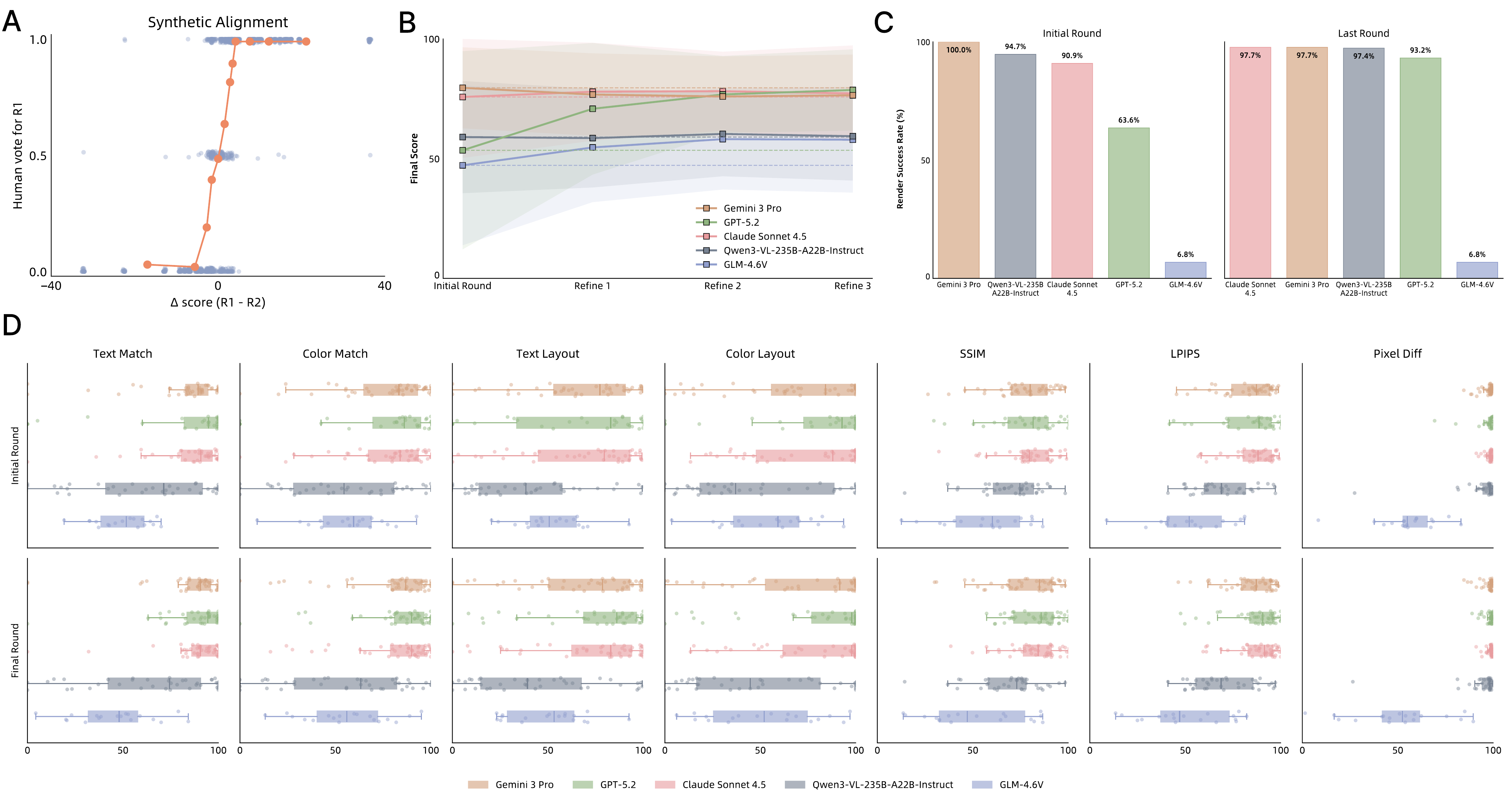}
    \caption{(A) Synthetic preference data relating the score difference between R1 and R2 to human preference, shown as jittered scatter and binned means. (B) Multi-round final score trends with mean curves and \(\pm\)1 standard deviation bands. (C) Render success rates for initial and final rounds shown as paired bar charts. (D) Boxplots of metric breakdowns for initial and final rounds.}
    \label{fig:Benchmarking}
\end{figure*}

Table~\ref{tab:model-performance} summarizes single-round and multi-round results using the metrics in Section~\ref{sec:metrics}. For multi-round generation, $\bar{S}$ and RSR are computed on the final round. In single-round generation, Gemini 3 Pro achieves the best FinalScore of 79.6 with 100.0\% RSR, while Claude Sonnet 4.5 attains the highest similarity with $\bar{S}=81.4$. In multi-round generation, iterative editing generally improves RSR, most notably for GPT-5.2 from 63.6\% to 93.2, and yields the largest similarity gain for GPT-5.2 from 78.3 to 84.9. Claude Sonnet 4.5 achieves the best multi-round FinalScore of 80.4, followed by Gemini 3 Pro at 79.5 and GPT-5.2 at 79.1.

Figure~\ref{fig:Benchmarking}B-D shows that multi round interaction improves both rendering success rates and similarity scores in most cases. 

\begin{table*}[t]
  \centering
  \small
  \renewcommand{\arraystretch}{1.12}
  \setlength{\tabcolsep}{2.5pt}

  \begin{adjustbox}{max width=\linewidth}
  \begin{tabularx}{\linewidth}{L{5.8cm} Y Y Y Y Y Y}
    \toprule
    \multirow{2}{*}{\textbf{Model}} &
    \multicolumn{3}{c}{\textbf{Single-Round}} &
    \multicolumn{3}{c}{\textbf{Multi-Round}} \\
    \cmidrule(lr){2-4} \cmidrule(lr){5-7}
    & \textit{$\bar{S}$} & \textit{RSR} & \textit{FinalScore}
    & \textit{$\bar{S}$} & \textit{RSR} & \textit{FinalScore} \\
    \midrule
    GLM-4.6V & 54.0 & 6.8\%  & 3.7  & 77.0 & 6.8\%  & 5.2  \\
    Qwen3-VL-235B-A22B-Instruct & 62.3 & 94.7\% & 59.0 & 63.5 & 97.4\% & 61.9 \\
    Claude Sonnet 4.5 & \textbf{81.4} & 90.9\% & 74.0 & 82.2 & \textbf{97.7}\% & \textbf{80.4} \\
    GPT-5.2 & 78.3 & 63.6\% & 49.8 & \textbf{84.9} & 93.2\% & 79.1 \\
    Gemini 3 Pro & 79.6 & \textbf{100.0}\% & \textbf{79.6} & 81.3 & \textbf{97.7}\% & 79.5 \\
    \bottomrule
  \end{tabularx}
  \end{adjustbox}

  \caption{Model performance on single-round and multi-round generation.}
  \label{tab:model-performance}
\end{table*}

\section{Pilot Study: Synthetic Repair Data and Agentic RL}
\label{sec:exploratory_study}

\begin{figure*}[t]
    \centering
    \includegraphics[width=1\linewidth]{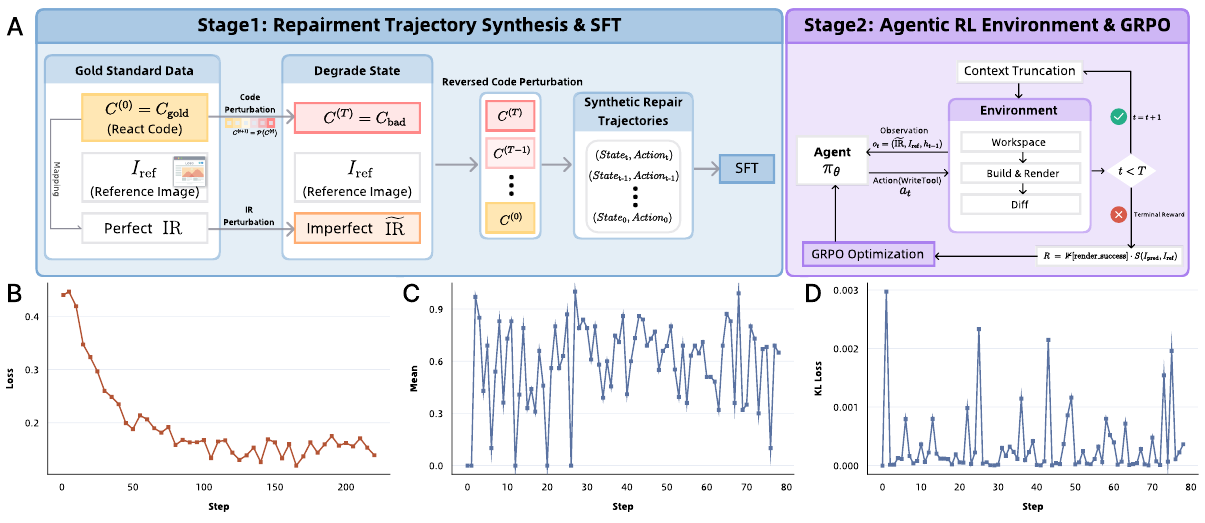}
    \caption{Overview of the pilot post-training study. (A) Synthetic trajectory construction for SFT and the segmented-rollout GRPO setup. (B) SFT training loss. (C) Mean similarity score during GRPO training. (D) KL divergence to the reference policy during GRPO training.}
    \label{fig:data_synthesis_and_training}
\end{figure*}

\subsection{Motivation}
\label{sec:motivation_iteration}

Under a fixed toolchain, each output is executable and can be rendered for direct comparison with the reference image (Section~\ref{sec:metrics}), which makes multi-round editing a natural alternative to single-pass generation. In 1D-Bench, the exported IR provides structural cues but is imperfect, so models may benefit from iterative revision using execution feedback. Since Section~\ref{sec:main_results} shows that multi-round editing improves final-round performance in most cases, we conduct a pilot post-training study under the same evaluation interface, with access to the IR, the reference image, and a persistent workspace via WriteTool.

\subsection{Synthetic Repair Trajectories}
\label{sec:synthetic_repair}

Figure~\ref{fig:data_synthesis_and_training}A summarizes the construction of synthetic trajectories for both initial generation and repair under the WriteTool-only interface. We first obtain high-quality reference workspaces $C_{\text{gold}}$ from two sources: high-scoring projects collected from an internal platform and projects synthesized by Qwen3-Coder-Plus ~\citep{qwen_qwen3_coder_next_tech_report}. For each $C_{\text{gold}}$, we render the workspace to produce the reference image $I_{\text{ref}}$, export a corresponding IR, and apply mild perturbations to produce an imperfect IR.

To create repair supervision, we derive a degraded workspace $C_{\text{bad}}$ by applying component-level code perturbations that mimic common errors, such as small numeric drifts in style values and localized structural changes in JSX. We render $C_{\text{bad}}$ to obtain $I_{\text{bad}}$, together with a similarity score and a diff heatmap relative to $I_{\text{ref}}$. From each pair $(C_{\text{gold}}, C_{\text{bad}})$, we generate two trajectories that share the same endpoint $C_{\text{gold}}$. The first is an initial-generation trajectory conditioned on $(\widetilde{\mathrm{IR}}, I_{\text{ref}})$ that writes the workspace through file overwrites. The second is a repair trajectory conditioned on $(\widetilde{\mathrm{IR}}, I_{\text{ref}}, C_{\text{bad}}, I_{\text{bad}}, \text{score}, \text{diff})$ that restores the perturbed files. The difficulty of repair is controlled by the perturbation types and their coverage.

\subsection{Post-training: SFT and Segmented-rollout GRPO}
\label{sec:post_training}

We adopt a two-stage post-training recipe for the multi-round interface of 1D-Bench. Stage 1 performs supervised fine-tuning on synthetic generation and repair traces. Inspired by ReSum~\citep{wu2025resum}, Stage 2 applies GRPO with segmented rollouts under context resetting.

\paragraph{Segmented rollouts.}
Each instance maintains a persistent workspace. A multi-round run is executed as $K$ segments. In segment $k$, the model observes
\[
o_k=\Phi\big(IR, I_{\mathrm{ref}}, C_k, f_{k-1}\big),
\]
where $C_k$ is the workspace state and $f_{k-1}$ is the previous segment feedback with $f_0=\varnothing$. Within the segment, the model emits a sequence of WriteTool actions
\[
u_{k,1:m_k}\sim \pi_\theta(\cdot\mid o_k),
\]
which deterministically updates the workspace,
\[
C_{k+1}=T\big(C_k,u_{k,1:m_k}\big).
\]
The harness then builds and renders $C_{k+1}$ to produce feedback $f_k$. Before segment $k+1$, the dialogue context is cleared while the workspace $C_{k+1}$ is kept. We set $K=3$ to match evaluation.

\paragraph{Terminal reward.}
A run yields a segmented trajectory $\tau=\{(o_k,u_{k,1:m_k},f_k)\}_{k=1}^{K}$. We use a terminal-only reward from the final segment,
\[
R(\tau)=\mathbb{1}[\mathrm{render\_success}]\; S\!\left(I_{\mathrm{pred}}, I_{\mathrm{ref}}\right)\in[0,1].
\]
and treat intermediate scores only as feedback.

\paragraph{Segmented GRPO with advantage broadcasting.}
For each instance we sample $G$ trajectories $\{\tau_g\}_{g=1}^{G}$ and compute $R_g=R(\tau_g)$. We form a trajectory-level advantage ~\citep{shao2024deepseekmathpushinglimitsmathematical}
\[
\hat{A}_g=\frac{R_g-\mathrm{mean}(\{R_1,\ldots,R_G\})}{\mathrm{std}(\{R_1,\ldots,R_G\})},
\]
and broadcast it to all segments in the same trajectory. Let $r_{g,k}(\theta)$ be the GRPO importance ratio for tokens generated in segment $k$ of trajectory $g$. The objective is
\begin{equation*}
J_{\mathrm{seg\text{-}GRPO}}(\theta)
=\mathbb{E}\!\left[
\frac{1}{GK}\sum_{g=1}^{G}\sum_{k=1}^{K}
\min\!\left(
r_{g,k}(\theta)\hat{A}_g,\,
\mathrm{clip}\!\big(r_{g,k}(\theta),\,1-\epsilon,\,1+\epsilon\big)\hat{A}_g
\right)
\right].
\end{equation*}

This formulation matches the evaluation protocol by resetting context between segments while learning from a single final-round score.

\subsection{Results}
\label{sec:limited_gains}

Figure~\ref{fig:data_synthesis_and_training}B shows the SFT loss decreasing and then plateauing, indicating successful fitting of synthetic trajectories (tool-call format and edit patterns). However, this does not necessarily translate to better 1D-Bench performance, since the supervision mainly encourages reproducing trajectory structure rather than improving multi-round strategies.

During GRPO, Figure~\ref{fig:data_synthesis_and_training}C shows that the mean similarity score fluctuates without a sustained upward trend, suggesting no consistent improvement. Figure~\ref{fig:data_synthesis_and_training}D reports near-zero KL divergence for most updates, implying limited deviation from the reference policy. Together, these results indicate weak and unstable learning.

A plausible explanation is that the learning signal is dominated by sparse terminal rewards and discontinuous failure modes. In addition, WriteTool actions are implemented as full-file overwrites, which introduces high-variance macro edits that make credit assignment across rounds difficult. These properties can reduce the effective signal-to-noise ratio of policy gradients, leading to unstable optimization and limited policy change.

\section{Conclusion}
\label{sec:conclusion}

We present 1D-Bench, a real-world benchmark derived from e-commerce development workflows for evaluating design-to-code under practical constraints. Each instance provides both a reference rendering as the target, and an exported, potentially noisy IR as structural cues. It requires generating an executable React codebase under a fixed toolchain with an explicit component hierarchy. We further standardize a multi-round setting where models iteratively apply component-level edits using execution feedback.

Across commercial and open-weight MLLMs, we find that multi-round generation generally increases final performance. Finally, we conduct a pilot study for iterative D2C using synthetic repair trajectories and agentic GRPO. While SFT fits synthetic edit patterns, RL yields limited improvements, highlighting challenges from sparse terminal rewards and high-variance file-level updates. We hope 1D-Bench enables more comparable progress and motivates learning methods better aligned with iterative, executable front-end generation.

\bibliography{colm2026_conference}
\bibliographystyle{colm2026_conference}

\appendix

\section{Implementation of Data Construction}
\label{sec:Implementation of Data Construction}
We implement data construction as a two-stage pipeline. We first perform automatic cleaning to remove invalid, non UI, privacy risk, and near duplicate records. We then stratify the cleaned pool to form a balanced evaluation subset. Each raw record contains a reference rendering treated as reference target paired with an exported structural IR that may be imperfect.
\subsection{Automatic cleaning}
We drop records with missing or unreadable artifacts. The remaining records are filtered in a fixed order:
\begin{enumerate}
  \item aspect ratio in \([1/4.5, 4.5]\)
  \item near-duplicate removal using CLIP (ViT-B-32), remove if cosine similarity \(\ge 0.95\)
  \item transparent background ratio \(\le 0.1\)
  \item low-variation images, remove if \(\texttt{std\_mean}<10\) and \((\texttt{unique\_ratio}<0.05 \text{ or } \texttt{entropy}<1.5)\)
  \item VLM full-page screen with decision in keep, remove, and review
  \item VLM poster or illustration screen with decision in keep, remove, and review
  \item privacy compliance via OCR plus pattern-based PII detection, followed by a VLM privacy screen
\end{enumerate}
All \texttt{review} cases are manually inspected and we retain only records confirmed to be UI relevant and safe.

\subsection{IR quality labeling and stratified sampling}
We compute IR statistics and define a coarse complexity label by the 33rd/66th percentiles of node counts (easy/mid/hard). We then conduct a manual review to confirm and correct the complexity label for each task. We estimate IR reliability by rendering the exported IR in a deterministic HTML renderer, capturing an IR-based screenshot with a headless browser, and comparing it against the reference image. We label IR quality as \texttt{perfect} if similarity is \(\ge 0.95\) and \texttt{imperfect} otherwise. Platform (\texttt{desktop}/\texttt{mobile}/\texttt{component}) and primary language (\texttt{Chinese}/\texttt{English}/\texttt{other}) are labeled by a VLM. All VLM-based screening and labeling steps use a locally deployed Qwen3-VL-30B-A3B-Instruct model running on 4 H20 GPUs.

\subsection{Filtering prompts}
\begin{figure*}
    \centering
    \includegraphics[width=1\linewidth]{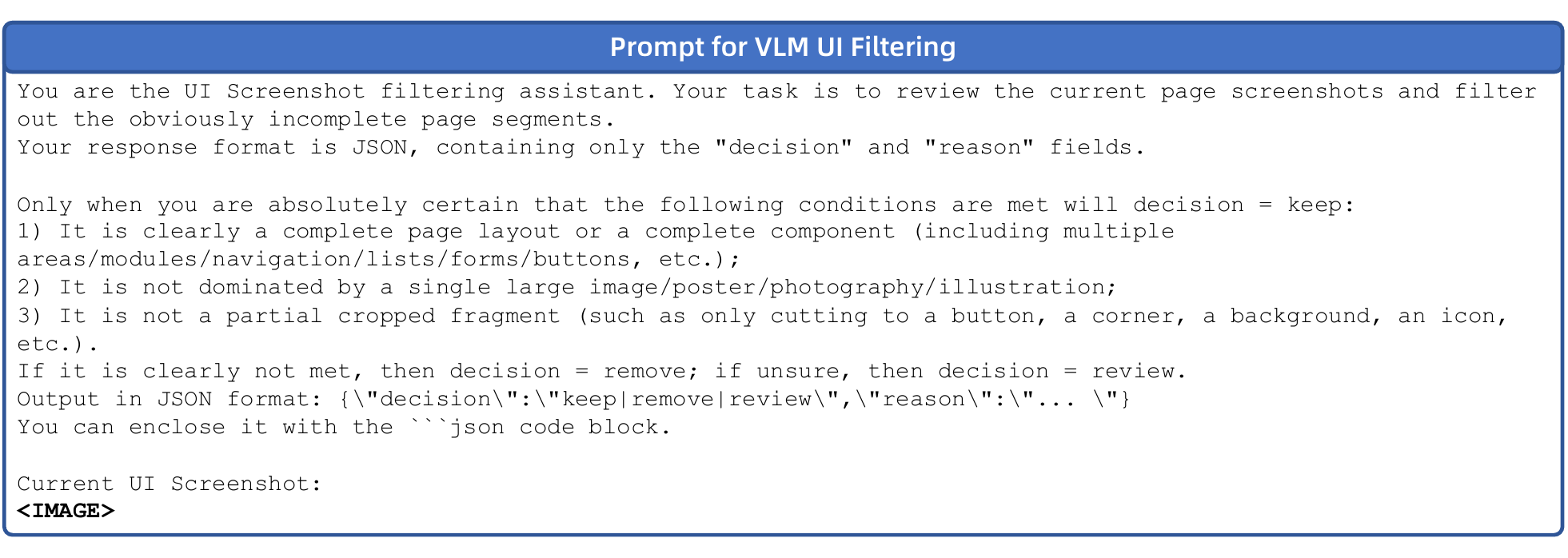}
    \includegraphics[width=1\linewidth]{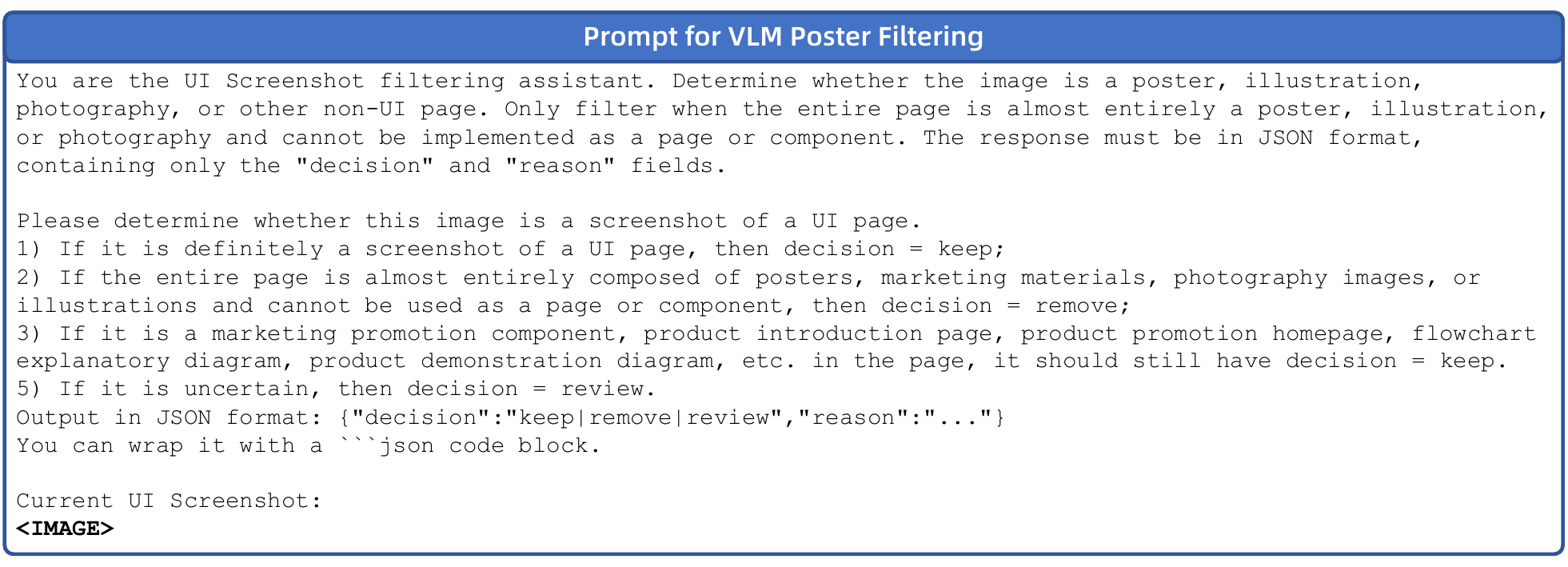}
    \includegraphics[width=1\linewidth]{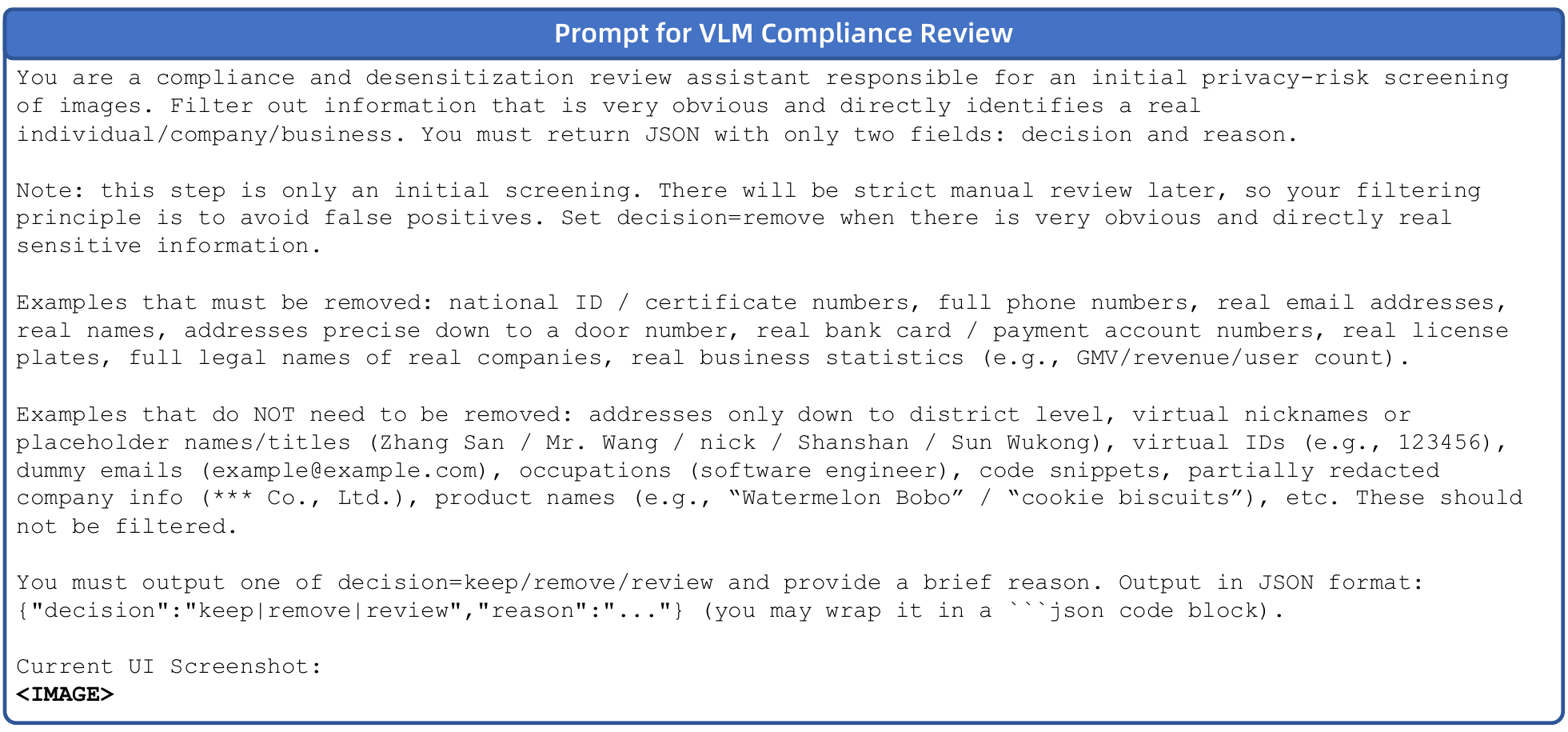}
    \caption{Prompts for VLM UI check}
    \label{fig:Prompt for VLM UI check}
\end{figure*}

We use a small set of VLM screening prompts to filter non page samples and obvious privacy risks (Figure~\ref{fig:Prompt for VLM UI check}). Each prompt requests a minimal JSON object with fields \texttt{decision} and \texttt{reason}. The \texttt{decision} is one of keep, remove, review. Review cases are sent to manual inspection. All screening is performed by the locally deployed Qwen3-VL-30B-A3B-Instruct model on 4 H20 GPUs, and no data are transmitted outside the local environment.

\section{Implementation of Metrics}
\label{sec:Implementation of Metrics}

We adopt an execution-based evaluation protocol. For each generated React project, we build and render it to obtain a deterministic screenshot. We then compute a similarity score against the reference image. The scoring function follows the composite design deployed in our production evaluation to ensure stable and reproducible measurements.

\paragraph{Rendering success rate.}
We treat executability as a first-class metric. A run is successful if the harness produces a nonempty screenshot within a fixed timeout. Otherwise the run is a rendering failure. We denote the rendering success rate by \textit{RSR}. The harness performs the following steps.
\begin{enumerate}
  \item \textbf{Deterministic build}. Clean stale build artifacts, install dependencies under a fixed toolchain, and build the project with a timeout.
  \item \textbf{Static serving}. Serve the built static assets via a lightweight HTTP server on a randomly selected free port.
  \item \textbf{Headless screenshot}. Use Playwright with a Chromium backend to capture a screenshot of the root container. We wait for \texttt{networkidle}, then resize the browser viewport to the tight bounding box of root container and its descendants to reduce cropping caused by fixed or absolute positioning.
\end{enumerate}
We retry screenshot capture up to three times. The rendering success rate is the fraction of tasks that are successful.

\paragraph{Visual similarity normalization.}
Given the rendered screenshot \(I_{\text{pred}}\) and reference image \(I_{\text{ref}}\), the scorer outputs a similarity \(S\in[0,1]\). We also generate a diff heatmap that visualizes pixel level discrepancies. The scoring pipeline is modular. If OCR, SSIM, or LPIPS cannot be computed, the corresponding term is omitted and the remaining weights are renormalized over the available signals.

\paragraph{Image preprocessing and size alignment.}
Before detection and matching, we apply transparency-aware preprocessing. If an image has an alpha channel, pixels with \(\alpha < \tau_k\) are treated as background and set to white in grayscale space. For transparent images, we evaluate multiple alpha thresholds \(\tau_k \in \{20,40,60,80,100,150\}\) and merge detections across thresholds; for non-transparent/general processing, we use a default alpha threshold of 200. For very dark foreground regions, we apply brightness amplification and contrast stretching. To avoid upscaling, we set \(w=\min(w_{\text{ref}}, w_{\text{pred}})\) and \(h=\min(h_{\text{ref}}, h_{\text{pred}})\), and resize both images to \((w,h)\). We run all detectors on these aligned images.

\paragraph{Element-level completeness for text blocks.}
We detect text blocks in both images using an OCR pipeline that combines a DBNet text detector and a CRNN recognizer. We set \texttt{max\_side\_len=960} and a detection score threshold of \texttt{0.6}. Each detection yields a bounding box and a recognized string.
We match each reference text block to at most one screenshot text block using a weighted similarity:
\[
s_{\text{text}} = 0.6\, s_{\text{content}} + 0.3\, s_{\text{pos}} + 0.1\, s_{\text{size}},
\]
where \(s_{\text{content}}\) is a normalized edit distance similarity over lowercased strings, \(s_{\text{pos}}\) is a center distance similarity normalized by the image diagonal, and \(s_{\text{size}}\) is an area ratio similarity. We accept matches when \(s_{\text{text}}\ge 0.5\) and resolve spatial conflicts to enforce a one to one assignment. We avoid hard binarization by mapping the match confidence through a piecewise saturation function \(f(\cdot)\), and we average \(f(s_{\text{text}})\) over all reference text blocks. Unmatched blocks contribute zero.

\paragraph{Element-level completeness for nontext blocks.}
To capture nontext visual regions, we detect blocks using multi scale Canny edge detection followed by contour extraction and bounding boxes. For transparent images, we use multiple Canny thresholds and exposure or gamma sweeps, then merge edge maps by pixelwise maximum and apply morphological closing. Candidate blocks are filtered by area, overlap is suppressed using non maximum suppression, and blocks overlapping OCR regions are removed to avoid double counting. For each reference block, we crop a grayscale template from the reference image and search it in the predicted image using normalized cross correlation template matching. We use a match threshold around \(0.5\), and we additionally check the original location with a confidence threshold of \(\ge 0.8\). As with text, we enforce a one to one assignment with spatial conflict handling, and we aggregate block completeness by averaging the saturated confidence.

\paragraph{Layout score.}
Presence alone can be insufficient when elements drift. Let \(c(b)\) denote the center of a box \(b\), and let \(\delta(b)\) denote its diagonal length. We define the layout similarity as
\[
\begin{aligned}
c_i &= c(b_i), \\
\bar{\delta} &= \tfrac{1}{2}\big(\delta(b_1)+\delta(b_2)\big), \\
s_{\text{layout}}(b_1,b_2) &= \max\left(0, 1-\frac{d(c_1,c_2)}{\bar{\delta}}\right).
\end{aligned}
\]
We average these scores over reference text blocks and over reference nontext blocks. When both types are present, we combine them with weights \(0.6\) for text and \(0.4\) for nontext. Otherwise the available type receives weight \(1.0\).

\paragraph{Perceptual image similarity.}
In addition to element cues, we compute image level similarity signals, each mapped to \([0,1]\). These include SSIM, a normalized pixel difference score based on MSE, MAE, or RMSE, and LPIPS. LPIPS distances \(d\) are converted to similarity by \(\exp(-d)\). All signals reuse the same preprocessing to improve robustness under transparency. We use weights \(w_{\text{lpips}}=0.8\), \(w_{\text{ssim}}=0.1\), and \(w_{\text{pix}}=0.1\), and renormalize them over the available methods.

\paragraph{Composite score and final reporting.}
The final similarity score is a weighted combination:
\[
S = 0.5\, S_{\text{img}} + 0.3\, S_{\text{comp}} + 0.2\, S_{\text{layout}},
\]
where \(S_{\text{comp}}\) averages text and nontext completeness when both exist, and otherwise uses the available type. If neither text nor nontext blocks are detected, we set \(S=S_{\text{img}}\). As defined in Section~\ref{sec:metrics}, we report \(\bar{S}\) over successfully rendered instances and the rendering success rate \textit{RSR}. The summary score is \(\textit{FinalScore}=\bar{S}\cdot \textit{RSR}\).

\paragraph{Preference-based calibration.}
To validate the automated score against human judgments, we run a pairwise preference study on synthetic perturbations. Starting from an IR to HTML rendering, we generate two perturbed variants, denoted R1 and R2, using controlled DOM and style edits such as sibling swaps, node moves, and numeric CSS drifts. We render both variants and score each against the same reference image. Annotators view the reference and the two variants side by side and select whether R1 or R2 is closer, or choose uncertain. With multiple annotators, we compute majority vote and consensus. We report agreement accuracy between the metric winner and the human majority, a calibration curve that relates \(P(\text{human picks R1})\) to \(\Delta \text{score}=\text{score(R1)}-\text{score(R2)}\), and disagreement as a function of \(|\Delta \text{score}|\).

\begin{figure*}
    \centering
    \includegraphics[width=1\linewidth]{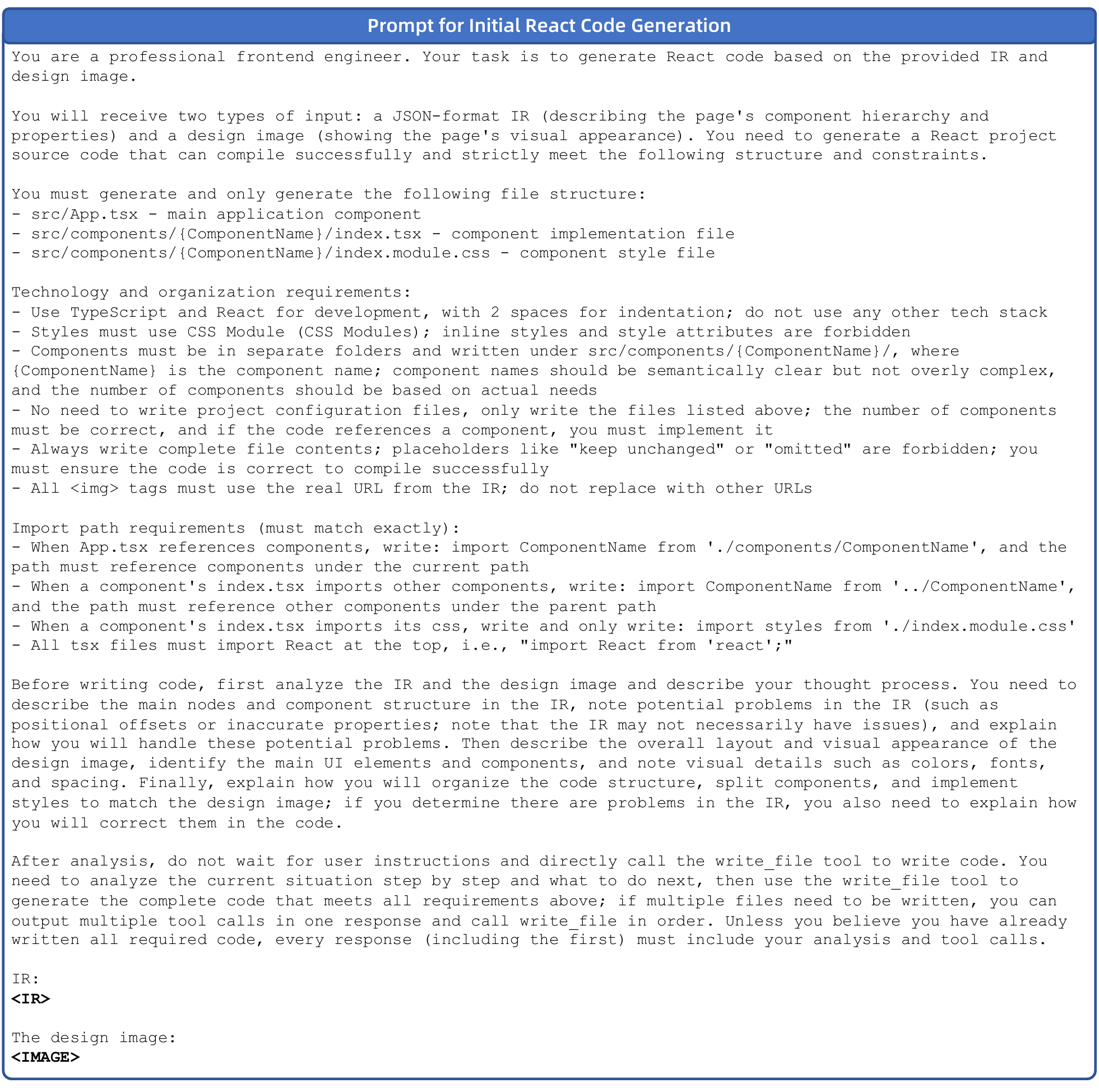}
    \caption{Prompt for Initial React Code Generation}
    \label{fig:Prompt for Initial React Code Generation}
\end{figure*}

\begin{figure*}
    \centering
    \includegraphics[width=1\linewidth]{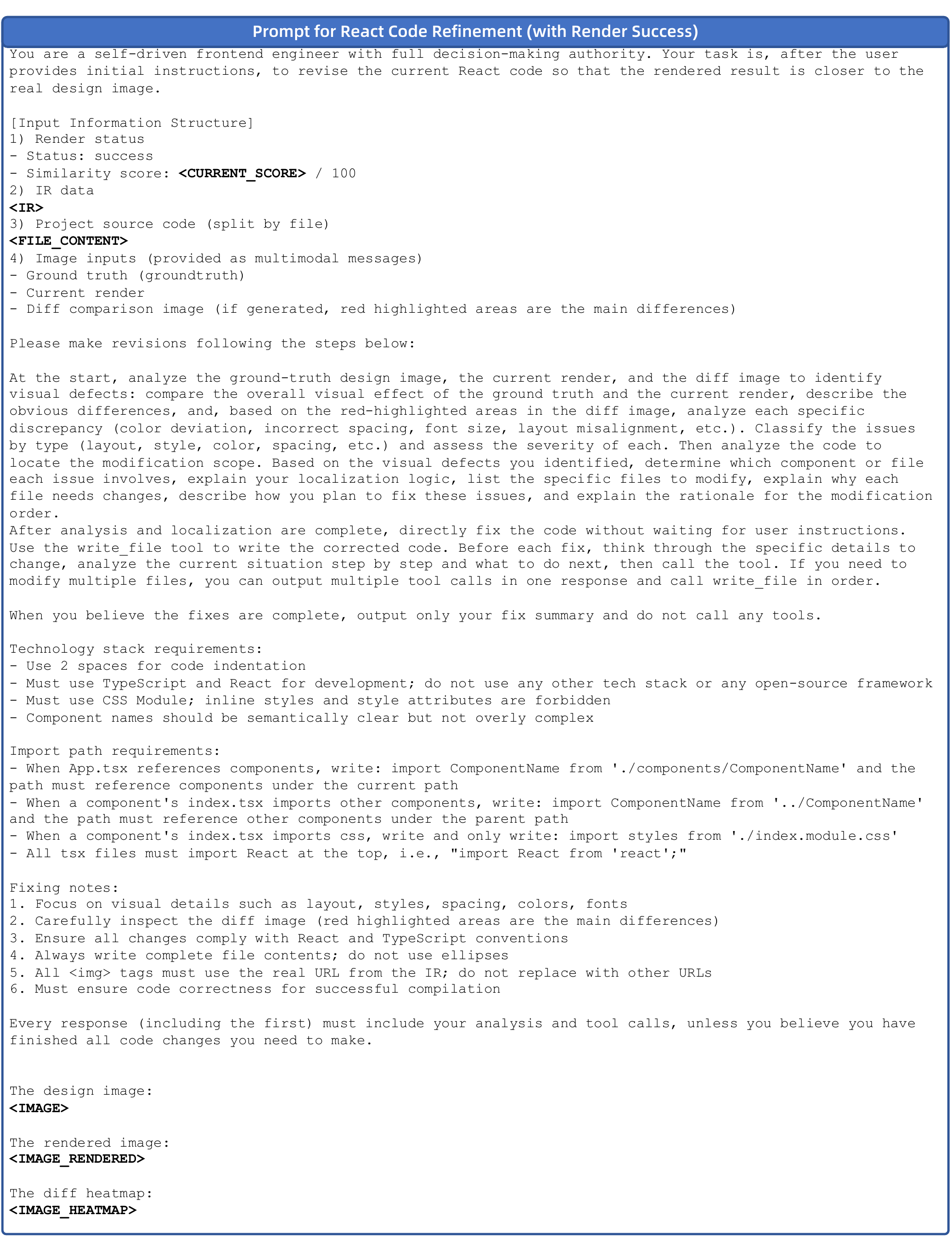}
    \caption{Prompt for React Code Refinement (with Render Success)}
    \label{fig:Prompt for React Code Refinement (with Render Success)}
\end{figure*}

\begin{figure*}
    \centering
    \includegraphics[width=1\linewidth]{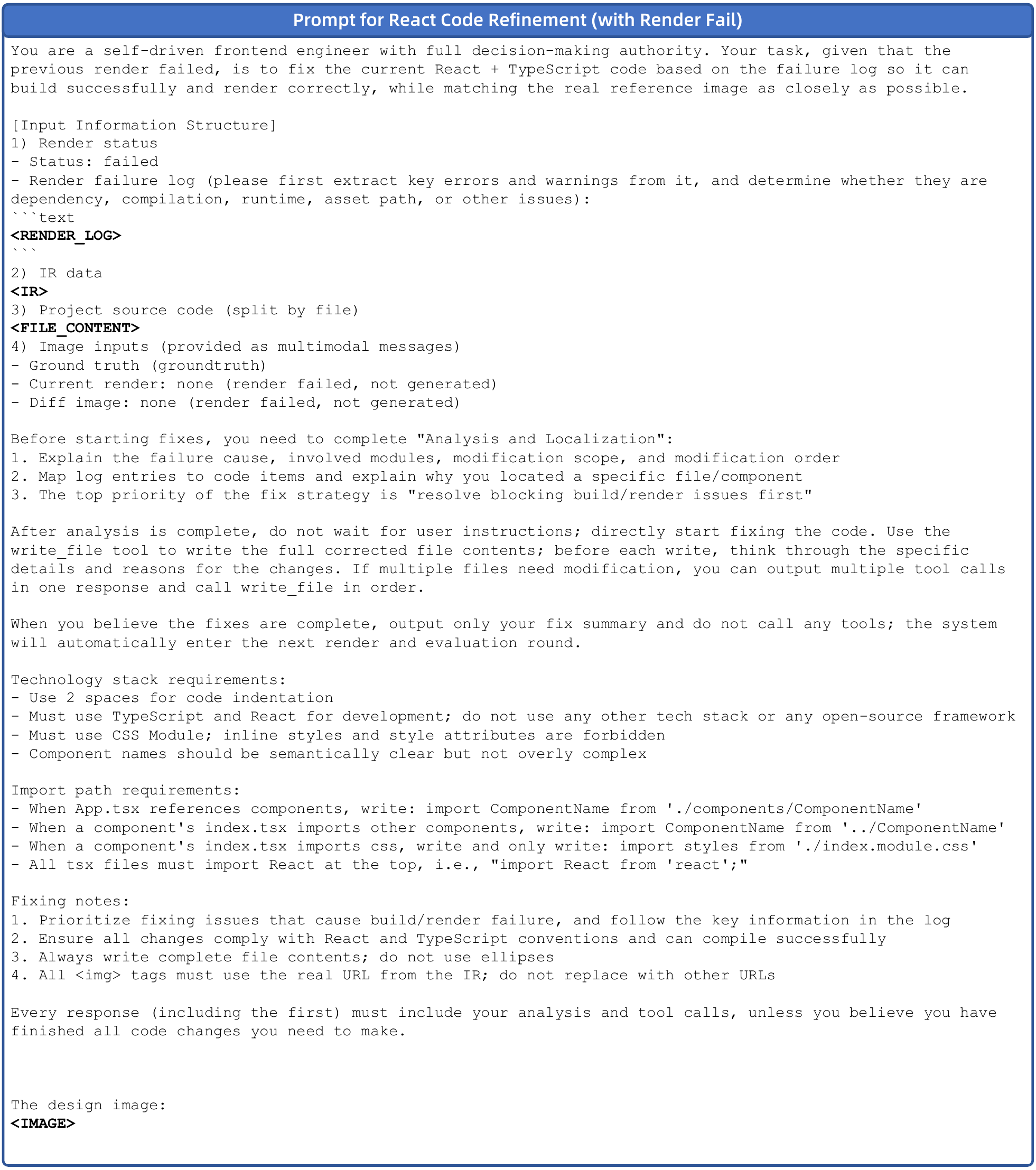}
    \caption{Prompt for React Code Refinement (with Render Fail)}
    \label{fig:Prompt for React Code Refinement (with Render Fail)}
\end{figure*}

\section{Implementation of Benchmarking}
\label{sec:Implementation of Benchmarking}

We implement benchmarking as an execution-based harness that evaluates models by the quality of a runnable React project rather than a single static snippet. The harness supports both one-shot generation and multi-round refinement under identical toolchain and rendering settings.

\paragraph{Fixed scaffold and controlled execution.}
For every instance, we start from the same dependency-locked React scaffold and build configuration. The harness installs dependencies, builds the project, and renders the resulting page in a headless browser with a fixed viewport and timeouts. This controls evaluation variance due to engineering choices and ensures comparability across models.

\paragraph{WriteTool-only interface.}
Models interact with the workspace through a single deterministic WriteTool API that can create or overwrite files. We support OpenAI-style structured tool calling when available, and otherwise parse tool-call payloads from plain-text outputs to remain compatible with servers that do not expose structured tool calls. When a model emits multiple writes to the same file within one turn, we execute only the last write to make action semantics well-defined.

\paragraph{Single-round generation.}
In the single-round setting, the model receives the exported IR together with the reference image and writes the full project in one pass. The harness then builds and renders the project and computes the visual similarity score and rendering success indicator (Section~\ref{sec:metrics}). The prompt is in Figure~\ref{fig:Prompt for Initial React Code Generation}.

\paragraph{Multi-round refinement with visual feedback.}
In the multi-round setting, we iterate up to a fixed number of rounds. Each round begins by building and rendering the current workspace. If rendering succeeds, we compute (i) the scalar similarity score and (ii) a pixel-level diff visualization between the current render and the reference. If rendering fails, we return the build/runtime logs. The next model call is conditioned on the original inputs, the current code snapshot, and the round feedback. We stop early when the score reaches a configurable target, otherwise we continue until the maximum rounds. The prompt is in Figure~\ref{fig:Prompt for React Code Refinement (with Render Success)} and Figure~\ref{fig:Prompt for React Code Refinement (with Render Fail)}

\section{Agentic RL Implementation}
\label{sec:Agentic RL Implementation}
We implement agentic RL on top of the ROLL framework with GRPO as the advantage estimator and a GEM-style environment interface~\citep{wang2025reinforcement}. Ray is used to coordinate dataset access and environment workers, while the scheduler/LLM proxy in ROLL handles batched generation. The D2C environment exposes a step-wise interaction protocol in which one \emph{round} of editing (potentially containing multiple tool calls) is treated as a single RL step; inner tool invocations are micro-steps with zero reward, and the round reward is assigned after termination.

Each episode samples a task from a Ray-backed global dataset and initializes an isolated React workspace by copying a fixed scaffold. The environment returns the reference design image as the primary observation and provides the IR and metadata in the info dict. At the end of each round, the project is built and rendered via a capture script, then scored by an internal diff-based scorer that produces a percentage similarity and an optional heatmap. The scalar reward is the absolute score normalized to $[0,1]$; rendering or scoring failures yield zero reward while preserving failure logs for feedback. A target score can trigger early termination.

\section{Case Studies}
\label{sec:Case Study}
\begin{figure*}
    \centering
    \includegraphics[width=1\linewidth]{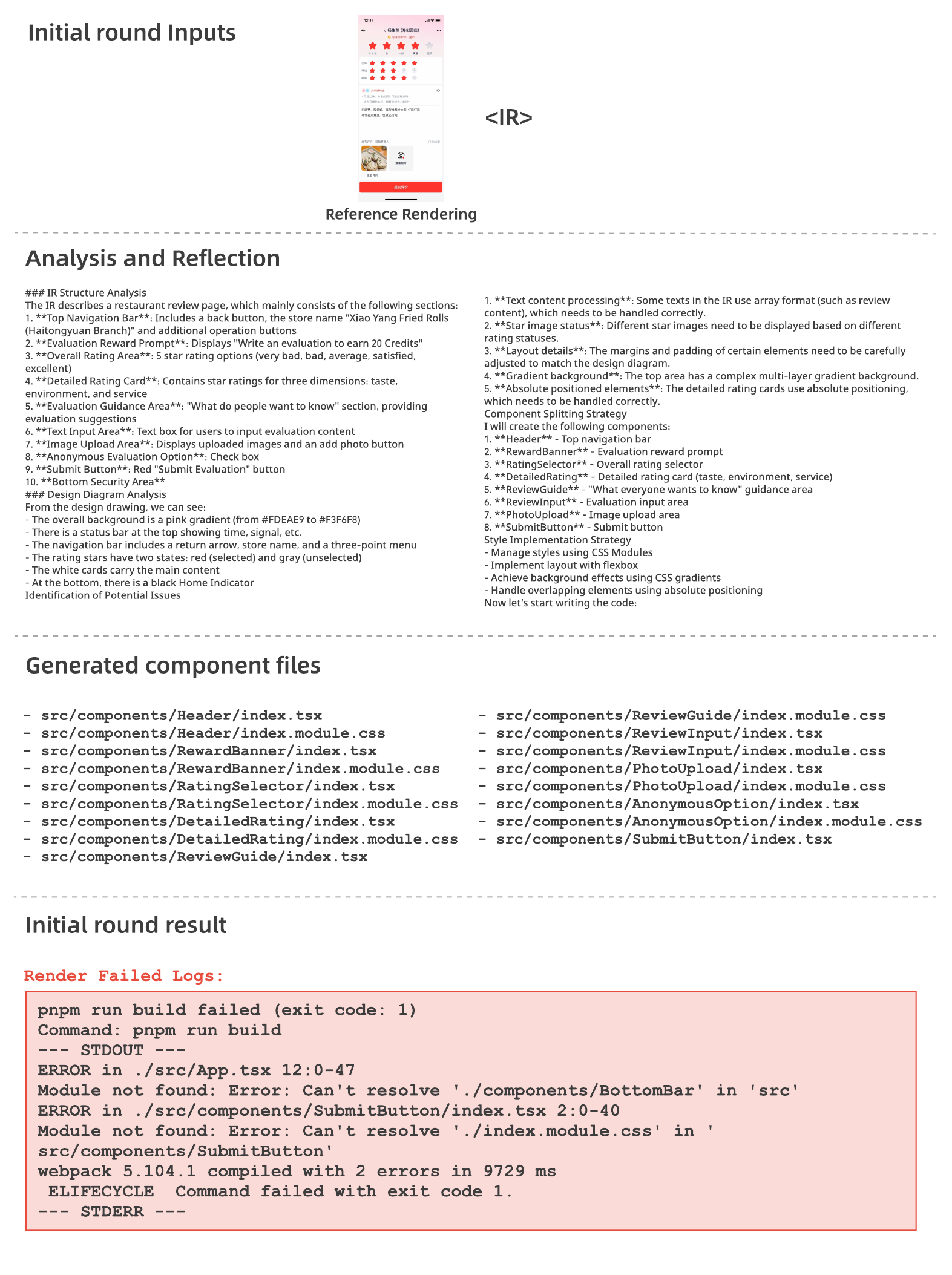}
    \caption{A Case of Initial Round}
    \label{fig:Case of Initial Round}
\end{figure*}

\begin{figure*}
    \centering
    \includegraphics[width=1\linewidth]{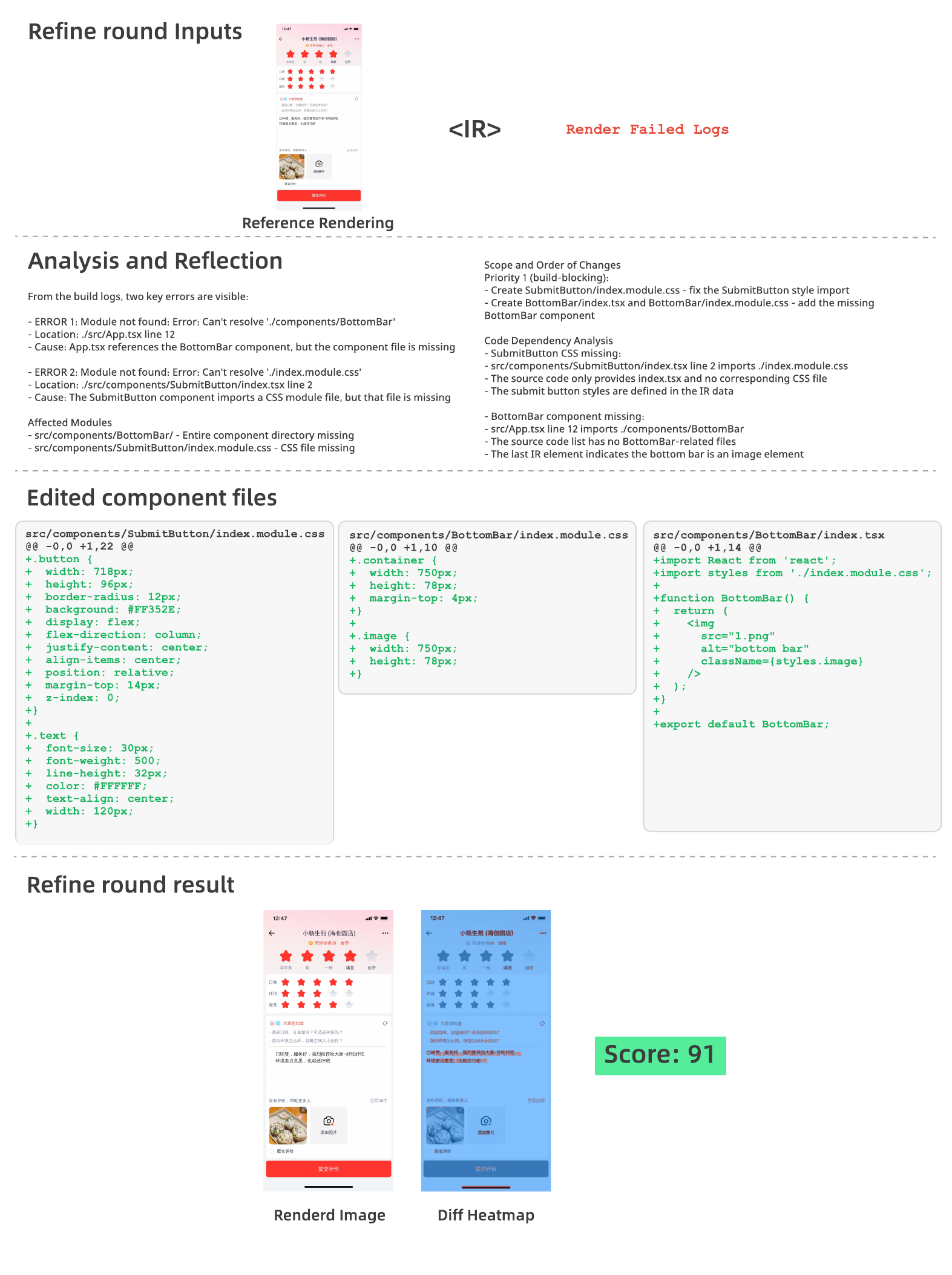}
    \caption{A Case of Refine Round}
    \label{fig:Case of Refine Round}
\end{figure*}

Refer to Figures~\ref{fig:Case of Initial Round} and~\ref{fig:Case of Refine Round} for details on the Claude 4.5 Sonnet model.

\end{document}